\documentclass[fleqn,usenatbib]{mnras}
 
\usepackage{amsmath}
\usepackage{multirow}
\usepackage{multicol}
\usepackage{graphicx}
\usepackage{newtxtext,newtxmath}
\usepackage{gensymb}
\title[The White Dwarf Mass-Orbital Period Relation]{The White Dwarf Mass-Orbital Period Relation Under Wind Mass Loss}

\author[S.-J. Gao \& X.-D. Li]{
	Shi-Jie Gao$^{1,~2}$\thanks{E-mail: gaosj@smail.nju.edu.cn} and
	Xiang-Dong Li$^{1,~2}$\thanks{E-mail: lixd@nju.edu.cn}
	\\
	$^1$School of Astronomy and Space Science, Nanjing University, Nanjing, 210023, People's Republic of China\\
	$^2$Key Laboratory of Modern Astronomy and Astrophysics, Nanjing University, Ministry of Education, Nanjing, 210023, People's Republic of China
}

\date{Accepted XXX. Received YYY; in original form ZZZ}
\pubyear{\today}

\begin{document}
\label{firstpage}
\pagerange{\pageref{firstpage}--\pageref{lastpage}}
\maketitle

\begin{abstract}
Helium white dwarfs (HeWDs) are thought to form from low-mass red giant stars experiencing binary interaction. Because the helium core mass of a red giant star is closely related to the stellar radius,  there exists well-known relation between the orbital period ($P_{\rm orb}$) and the mass ($M_{\rm WD}$) of the HeWDs, which is {almost} independent of the type of the companion star. Traditional derivation of the $M_{\rm WD}$--$P_{\rm orb}$ relation generally neglected the effect of wind mass loss from the red giants, while observations show that wind mass loss from red giants in binary systems is systematically higher than that from isolated stars. In this work, we calculate binary evolution with tidally enhanced stellar wind (TEW) and find that it causes significantly scatter of the traditional $M_{\rm WD}$--$P_{\rm orb}$ relation. The TEW can prevent the red giants from overflowing their Roche lobes and slow down the growth of the helium core, leaving a lower-mass HeWD for given orbital period. This scenario may account for some of the HeWD binaries that deviate from the traditional $M_{\rm WD}$--$P_{\rm orb}$ relation. However, we point out that observations of more HeWD binaries in wide orbits are needed to test the TEW model and to constrain the enhanced wind factor.
\end{abstract}


\section{Introduction}\label{sec:intro}

Single {low and intermediate mass} stars generally evolve to be carbon-oxygen white dwarfs (WDs). However, stars with masses $\lesssim2.3~{\rm M_\odot}$ in close binaries can lose their envelope through Roche-lobe overflow (RLOF) and finally leave helium white dwarfs (HeWDs). During the red giant branch (RGB), {the mass of the core} and the radius $R$ of the star are hardly dependent on the mass of the hydrogen-rich envelope\footnote{{The degenerate core is highly concentrated and the pressure drops dramatically within a thin burning shell above the core surface. The extended envelope is nearly weightless and has no influence on the burning shell. So the stellar luminosity, effective temperature and stellar radius are mainly dependent on core mass. See also Chapter 33 of \cite{Kippenhahn+2012}.}}\citep{Refsdal+1971,Kippenhahn+1981,Webbink+1983,deKool+1986,Han+1998,Kippenhahn+2012}. {On the other hand}, RLOF mass transfer requires that the donor radius $R$ is always close to its {Roche lobe} (RL) radius $R_{\rm L}$ which is determined by the orbital separation {and the mass ratio of the components} \citep{Eggleton+1983}. Consequently, there is specific correlation between the orbital periods $P_{\rm orb}$ and the masses $M_{\rm WD}$ of the HeWDs in the post-mass transfer binaries \citep{Joss+1987,Rappaport+1995,Tauris+1999}. 
Previous investigations showed that opacity, mixing length, convective overshooting and metallicity can influence the size of the RGB stars, and hence the $M_{\rm WD}$--$P_{\rm orb}$ relation \citep[e.g.,][]{Rappaport+1995,Tauris+1999,Chen+2013,Istrate+2016}, while the mass transfer efficiency, the masses of HeWD's progenitor and its companion, as well as the angular momentum loss mechanisms have insignificant effects on the $M_{\rm WD}$--$P_{\rm orb}$ relation \citep[e.g.,][]{DeVito+2010,Lin+2011,Chen+2013,Jia+2014}.

Observations of binary millisecond pulsars (MSPs) with HeWD companions are broadly consistent with the $M_{\rm WD}-P_{\rm orb}$ relation. However, some of the systems seem to have HeWDs {lighter} than expected at given $P_{\rm orb}$ \citep{Tauris+1999,Shao+2012}. Moreover, similar feature exists in other types of HeWD binaries with non-degenerate companions, e.g., main-sequence (MS) stars\footnote{{MS stars in MS+HeWD binaries were the accretors at the first mass transfer phase, during which the progenitor of the HeWD overflowed its RL. After the mass transfer, a rejuvenated MS star and a HeWD were left in the system. So, one can also expect the $M_{\rm WD}$--$P_{\rm orb}$ relation in MS+HeWD binaries.}} (see \autoref{sec:res}), indicating that the inconsistency has nothing to do with the nature of the companion stars, but should be related with the evolution of the HeWD progenitors.

Compared with mass loss via RLOF, wind mass loss from RGB stars is usually neglected in previous works when deriving the $M_{\rm WD}$--$P_{\rm orb}$ relation. However, {there is observational evidence indicating} that the wind loss rate of RGB stars is systematically higher in binary systems than that in single RGB stars \citep[e.g.,][]{Kenyon+1988,Seaquist+1993,Mikoajewska+2012}. {For example, \cite{Zamanov+2008} showed that the symbiotic giants rotate on average more than twice as fast as the field red giants, and as a result of the rapid rotation, symbiotic giants have on average $\sim 10$ times larger mass loss rates than normal red giants.} There is also independent evidence of the wind focusing towards the orbital plane in the symbiotic star EG~And \citep{Shagatova+2021}. The bipolarity of the recurrent nova {in the symbiotic star} RS~Ophiuchi may result from the enhanced wind in the equatorial disc \citep{Obrien+2006}. Some planetary nebulae, for example, the Southern Crab \citep{Corradi+2001}, show an asymmetric morphology which may result from the enhanced wind in the equatorial disc. {Observations of GX~1+4, the prototype of symbiotic X-ray binaries, indicate a mass loss rate $\sim 10^{-6}~{\rm M_\odot~yr^{-1}}$ of the giant star \citep{1998ApJ...497L..39C}, significantly higher than predicted by the mass-loss law of \citet{Reimers+1975}. Hence, it is necessary to explore the influence of wind mass loss on the evolution of RGB binaries and the resulting $M_{\rm WD}$--$P_{\rm orb}$ relation.}

This paper is organised as follows. In \autoref{sec:mothod}, we introduce the physical considerations and binary evolution model. We present our results in \autoref{sec:res} and compare with observations in \autoref{sec:com}. Finally, we {discuss our results and present our conclusions} in \autoref{sec:dis}.

\section{Methods and calculations }\label{sec:mothod}

\begin{figure*}
    \centering
    \includegraphics[width=\linewidth]{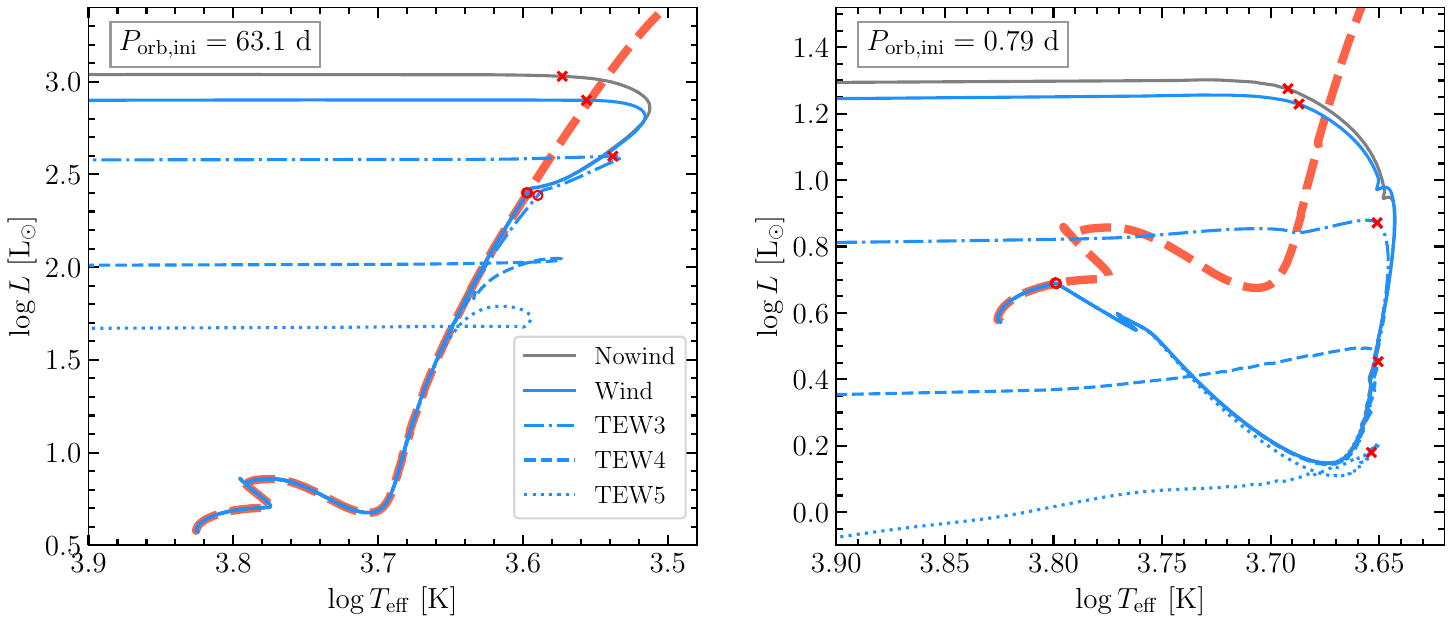}
    \caption{Hertzsprung-Russell Diagram of the donor in different models, with $(M_{\rm d,ini}$, $M_{\rm a,ini}$, $P_{\rm orb,ini})$=$(1.4~{\rm M_\odot}$, $1.5~{\rm M_\odot}$, $63.10~{\rm d})$ and $(1.4~{\rm M_\odot}$, $1.5~{\rm M_\odot}$, $0.79~{\rm d})$ for the left and right panels, respectively. Here, accretors are assumed to be NSs. The gray solid, blue solid, blue dash-dotted, blue dashed and blue dotted lines correspond to Models~\textit{NoWind, Wind, TEW3, TEW4 {\rm and} TEW5}, respectively. The red dashed line in each panel {depicts} the evolution of an isolated $1.4~{\rm M_\odot}$ star. The red circles and crosses denote the onset and end of RLOF, respectively.}
    \label{fig:HR}
\end{figure*}

\begin{figure*}
    \centering
    \includegraphics[width=\linewidth]{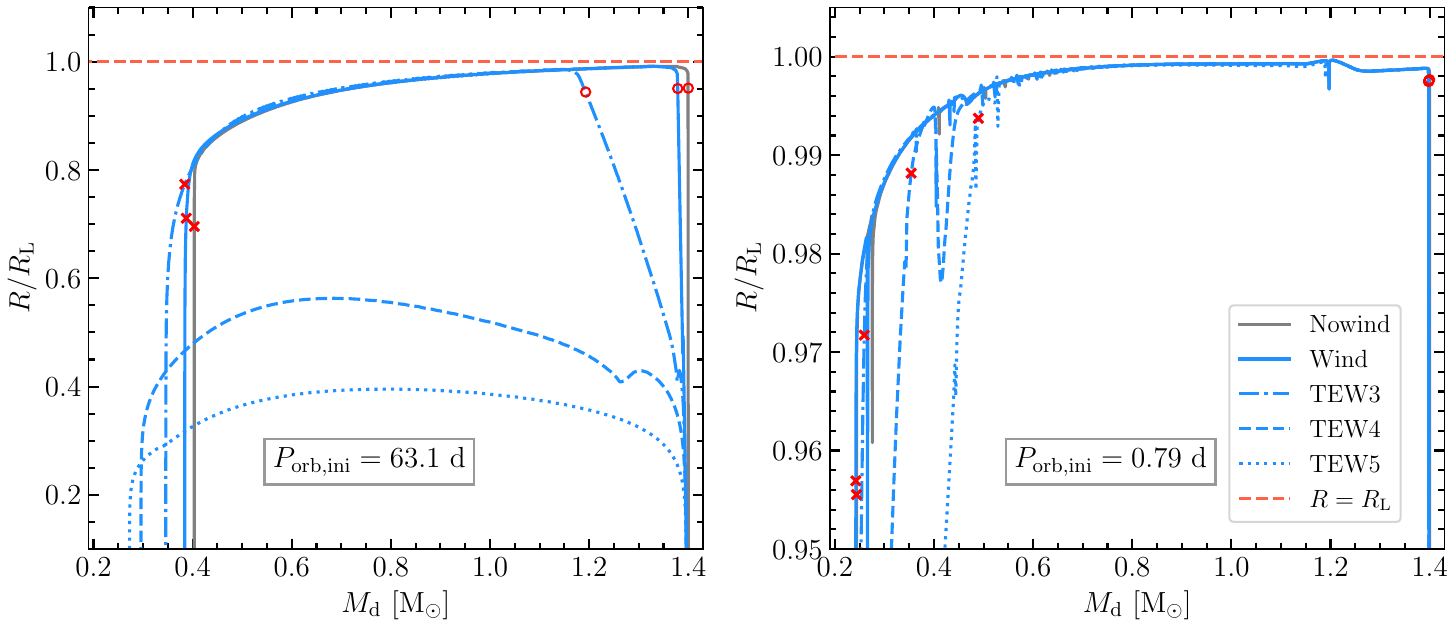}
    \caption{The ratio of the donor radius and its RL radius as a function of the {donor} mass. The red dashed line indicates $R=R_{\rm L}$. The red circles and crosses denote the onset and end of RLOF, respectively.}
    \label{fig:RRL}
\end{figure*}

\begin{figure*}
    \centering
    \includegraphics[width=\linewidth]{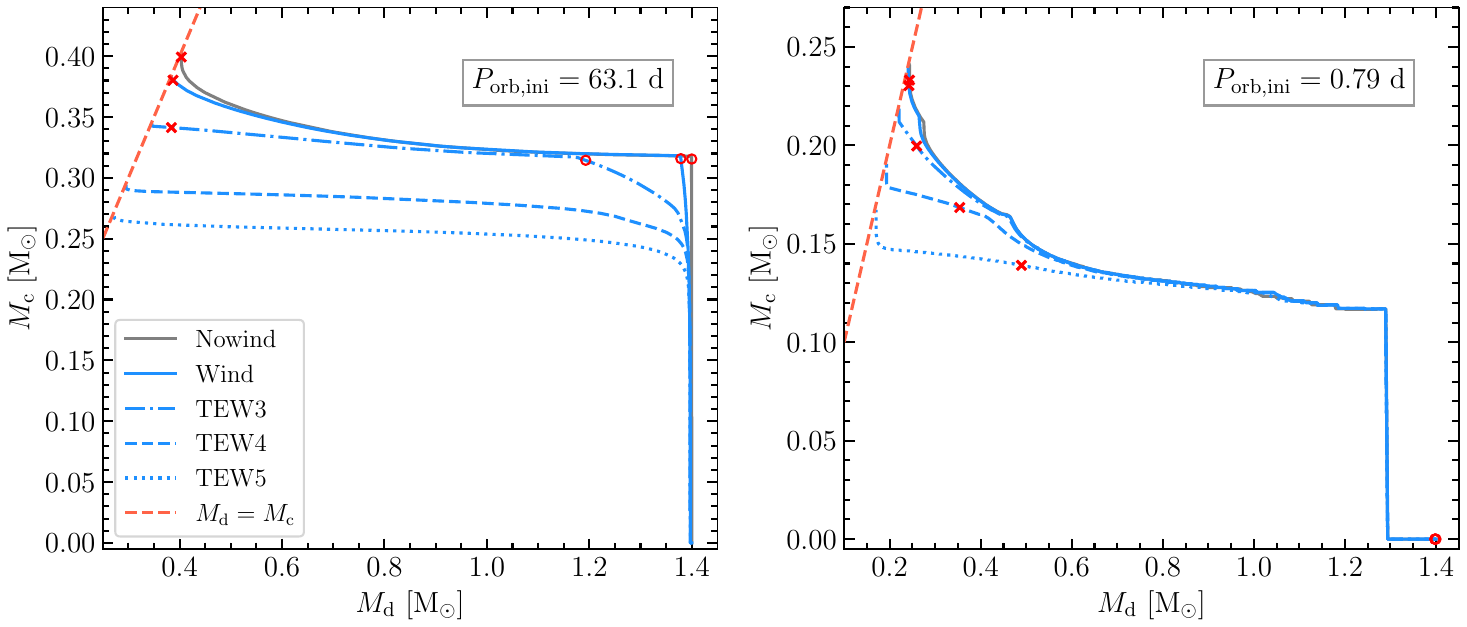}
    \caption{The helium core mass as a function of the {donor} mass. The red dashed line indicates $M_{\rm d}=M_{\rm c}$. The red circles and crosses denote the onset and end of RLOF, respectively.}
    \label{fig:MHeCore}
\end{figure*}

\begin{figure*}
    \centering
    \includegraphics[width=\linewidth]{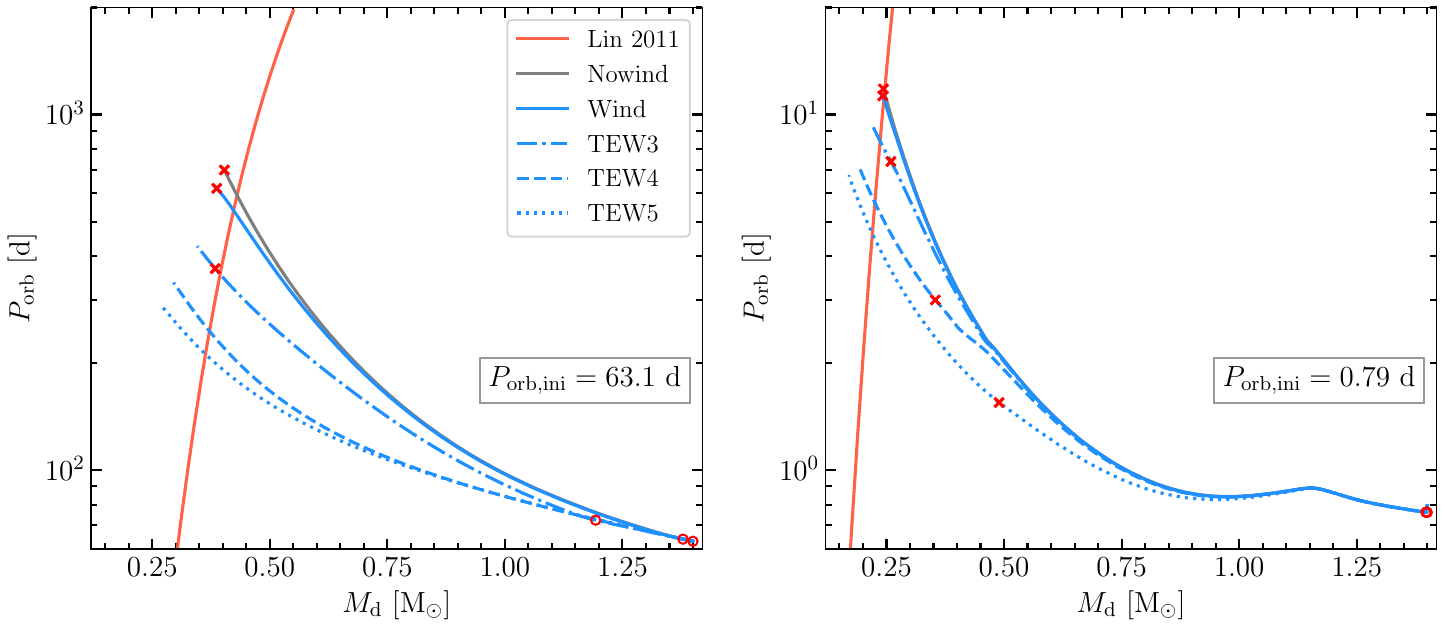}
    \caption{The orbital period as a function of the donor mass. The red solid line indicates the $M_{\rm WD}$--$P_{\rm orb}$ relations of \protect\cite{Lin+2011}. The red circles and crosses denote the onset and end of RLOF, respectively.}
    \label{fig:MPorb}
\end{figure*}

We use the state-of-the-art stellar evolution code \textsc{MESA}\footnote{\url{https://github.com/MESAHub/mesa}} \citep[version 22.11.1,][]{MESA+2011,MESA+2013,MESA+2015,MESA+2018,MESA+2019,MESA+2023} to evolve binary stars composed {by} a zero-age MS star (the progenitor of the HeWD/{the donor}) with mass $M_{\rm d}$ and a companion ({the accretor}) with mass $M_{\rm a}$. When the donor evolves to be a RGB star, we adopt the prescription of stellar wind loss rate from \cite{Reimers+1975}, i.e.,
\begin{equation}\label{eq:reimers}
        \dot M_{\rm w}=-4\times10^{-8}\times\eta_{\rm R}\left(\frac{M_{\rm d}}{1~\rm M_\odot}\right)^{-1}\left(\frac{L}{\rm 10^3L_\odot}\right)\left(\frac{R}{10^2\rm R_\odot}\right)~{\rm M_\odot~yr^{-1}},
\end{equation}
where $L$ and $R$ are the luminosity and radius of the donor star respectively, and $\eta_{\rm R}$ is a scaling factor and usually taken to be $0.5$. The Reimers' formula is valid when the donor is earlier than the asymptotic giant branch (AGB, or the central helium abundance by mass is $<10^{-4}$). Besides, \cite{Tout+1988} suggested that tidal interactions and/or magnetic activities in binaries may enhance the wind mass-loss rate in cool subgiants/giants by a factor,
\begin{equation}\label{eq:tew}
    \eta_{\rm T}=1+B_{\rm w}\times\min\left[\left(\frac{R}{R_{\rm L}}\right)^6,~\frac{1}{2^6}\right],
\end{equation}
{where $B_{\rm w}$ is an adjustable parameter and usually taken to be $\sim 10^3-10^4$, to be compatible with observations of RS~CVn binaries, barium and CH (methylidyne) stars, and Type~Ia supernovae \cite[e.g.,][]{Tout+1988,Han+1995,Chen+2011,Meng+2016,Wu+2021}.} The factor $(R/R_{\rm L})^6$ is from the tidal friction torque formula \citep{Zahn+1977,Campbell+1983}. {In the tidally enhanced stellar wind (TEW) model, the presence of a companion enhances the mass-loss rate regardless of its nature. The mechanism of the enhancement, although not physically understood, is rooted in tidal friction and the dynamo activity of the donor star. So it has also been adopted in researches on the evolution of interacting binaries containing a neutron star (NS) or black hole accretor \citep{1998ApJ...509..362R,2017MNRAS.464..237S}}. In our calculation, the TEW prescription is applied when the donor is on the RGB  with the central hydrogen abundance by mass $<1\times10^{-6}$. 

The mass transfer maintains through both capture of the stellar wind and RLOF.
The wind accretion efficiency $\beta_{\rm w}$ of the accretor is calculated with the Bondi-Hoyle-Lyttleton prescription \citep{Hoyle+1939,Bondi+1944,Bondi+1952,Edgar+2004}, {and the maximum efficiency is usually limited by $0.5$ to avoid the accretion rate of the accretor exceeding the wind mass-loss rate of the donor} \citep{MESA+2015}. The excess wind material (with a mass fraction $1-\beta_{\rm w}$) is assumed to leave the system taking the specific angular momentum of the donor \citep{MESA+2015}. 
For mass transfer via RLOF, we adopt the Kolb scheme \citep{Kolb+1990}. Different from the classical RLOF scheme in Eggleton’s \textsc{STARS} code \citep[e.g.,][]{Tauris+1999} which requires $R\geq R_{\rm L}$, here the mass transfer can proceed even when $R<R_{\rm L}$ because the donor stars usually have an extended atmosphere, and the pressure scale height at the stellar surface is taken into account to obtain the RLOF rate. The Kolb scheme leads to a lighter HeWD than with the classical RLOF scheme for a given orbital period \citep{Zhang+2021}. {We assume that half of the transferred mass via RLOF is accreted by the accretor and the other half is ejected from the binary, probably in the form of disk winds and outflows \citep[e.g.,][]{2002ApJ...565.1107P}, taking away the specific angular momentum of the binary.} {In the meantime, the accretion rate is limited by the Eddington accretion rate\footnote{{The Eddington accretion rate is defined as the accretion rate at which the radiation pressure from the accreted material can be balanced by the inward gravitational force \citep{Frank+2002}.}} ($\sim 4\times10^{-8}~{\rm M_\odot~yr^{-1}}$) for a $1.4~{\rm M_\odot}$ NS accretor or the thermal time-scale mass accretion rate ($\sim 10^{-7}~{\rm M_\odot~yr^{-1}})$ for a $1~{\rm M_\odot}$ MS accretor \citep{Hurley+2002+BSE}. We find that the $M_{\rm WD}$--$P_{\rm orb}$ relation is insensitive to the value of the mass transfer efficiency.}

We also take into account angular momentum loss caused by magnetic braking \citep{Rappaport+1983} and gravitation wave radiation \citep{Landau+1971}. Magnetic braking only works when the donor has a convective envelope and a radiative core simultaneously, and the braking index $\gamma_{\rm mb}$ is set to be $3.0$ \citep{MESA+2015}. The competition between mass transfer and magnetic braking results in a bifurcation period ($\sim 1-2~{\rm d}$)  which divides {binaries with open to tight orbits} \citep{Pylyser+1988}. The $M_{\rm WD}$--$P_{\rm orb}$ correlation is valid for diverging binaries with $P_{\rm orb}\gtrsim 2~{\rm d}$ and  there is large scatter in the HeWD masses when $P_{\rm orb}\sim 2~{\rm d}$ \citep{Istrate+2014}. For binaries with orbital periods $\gtrsim 10~{\rm d}$ mass loss dominates angular momentum loss, and both magnetic braking and gravitational wave radiation can be ignored.

\begin{table}
    \centering
    \setlength{\tabcolsep}{20pt}
    \caption{Models in our calculations.\label{tab:model}}
    \begin{tabular}{ccc}
        \hline
        Models&$\eta_{\rm R}$&$B_{\rm w}$\\
        \hline
        NoWind&$0$&$-$\\
        Wind&$0.5$&$0$\\
        TEW3&$0.5$&$1\times10^3$\\
        TEW4&$0.5$&$1\times10^4$\\
        TEW5&$0.5$&$1\times10^5$\\
        \hline
    \end{tabular}
\end{table}

\section{Results}\label{sec:res}

{We have evolved a grid of binaries consisting of an MS donor with  initial mass $M_{\rm d, ini}$ of $1.2$, $1.4$, $1.6$ and $1.8~{\rm M_\odot}$, and an NS or MS accretor with initial mass $M_{\rm a,ini}$ of $1.2$, $1.5$ and $1.8~{\rm M_\odot}$.} The lower mass limit of $M_{\rm d,ini}$ is taken to ensure that the MS lifetime is shorter than $12~{\rm Gyr}$. The initial orbital period $P_{\rm orb,ini}$ in logarithm is in the range of $[-0.5,~2.6]$ in steps of $0.1~{\rm dex}$ between $-0.5$ and 1.1, and $0.2~{\rm dex}$ between 1.2 and 2.6, respectively. {The orbits are assumed to be circular.} The metallicity of the donor star is assumed to be $Z=0.02$. In addition, we adopt the exponentially overshooting model for the stellar core with an efficiency factor $f_{\rm ov}=0.0016$ \citep{Herwig+2000}. We exclude binaries with RLOF at the beginning of the evolution, or with the donor star overflowing from the $L_{2}$ Lagrange point, or with the mass transfer rate via RLOF exceeding $1\times 10^{-4}~{\rm M_\odot~yr^{-1}}$. We only select evolutionary tracks leaving a HeWD remnant. We adopt the default definition in \textsc{MESA} of the boundary between the helium-rich core and the hydrogen-rich envelope to be the outermost location where the mass fraction of $^{1}{\rm H}\le 0.1$ and the mass fraction of $^4{\rm He}\ge 0.1$. To examine the effects of the TEW, we construct five models of stellar winds, that is, no stellar wind from the donor (\textit{NoWind}), normal wind with $B_{\rm w}=0$ (\textit{Wind}), TEW with $B_{\rm w}=10^3$ (\textit{TEW3}), $10^4$ (\textit{TEW4}), and $10^5$ (\textit{TEW5}). The parameters of the five models are summary in \autoref{tab:model}.

\autoref{fig:HR} shows the Hertzsprung-Russell Diagram of the donor in different models. Here, accretors are assumed to be NSs. In the left and right panels, $(M_{\rm d, ini}$, $M_{\rm a,ini}$, $P_{\rm orb,ini})=(1.4~{\rm M_\odot}$, $1.5~{\rm M_\odot}$, $63.10~{\rm d})$ and $(1.4~{\rm M_\odot}$, $1.5~{\rm M_\odot}$, $0.79~{\rm d})$, respectively. The gray solid, blue solid, blue dash-dotted, blue dashed and blue dotted lines correspond to Models~\textit{Nowind, Wind, TEW3, TEW4 {\rm and} TEW5}, respectively. The red circles and crosses denote the onset and end of RLOF, respectively. The red dashed lines {depicts} the evolution of an isolated $1.4~{\rm M_\odot}$ star ({with stellar parameters} as in Model~\textit{Wind}) for comparison. The maximum luminosity that the donor can reach in both cases decreases with increasing wind mass loss. Since the core mass of a RGB star increases with increasing  luminosity \citep[e.g.,][]{Joss+1987,Kippenhahn+2012}, a less massive HeWD is  left with larger $B_{\rm w}$.

\autoref{fig:RRL} shows the evolution of $R/R_{\rm L}$ as a function of the donor mass $M_{\rm d}$. The red dashed line indicates $R=R_{\rm L}$, and other line types and symbols have the same meaning as in \autoref{fig:HR}. The dips in the curves are mainly caused by the dredge-up process or hydrogen shell flashes \citep{Istrate+2014,Istrate+2016}. It is seen that the donor's radius does not always equal its RL radius (see the red dashed line). This is because in the Kolb scheme RLOF can occur even if $R$ does not exceed $R_{\rm L}$. In the left panel,  in Models~\textit{NoWind}, \textit{Wind} and \textit{TEW3}, RLOF starts when the donor is on the RGB, while in Models~\textit{TEW4} and \textit{TEW5} there is no RLOF because the wind mass loss drives the orbital evolution and the donor can not catch up the expansion of the RL. In the right panel, RLOF commences when the donor is on the late MS in all cases. It can be seen that the gap between $R$ and $R_{\rm L}$ becomes larger with increasing $B_{\rm w}$, leaving a lower-mass HeWD compared with the case of \textit{NoWind}. This can also be seen in \autoref{fig:MHeCore}, which shows the relation between the donor mass and the helium core mass ($M_{\rm c}$). Compared with Model~\textit{Nowind}, wind mass loss can significantly limits the core growth. Moreover, TEW can continue stripping the donor mass after the RLOF, 
resulting in a lower-mass HeWD. For example, when $(M_{\rm d,ini},$ $M_{\rm a,ini},$ $P_{\rm orb})=(1.4~{\rm M_\odot}$, $1.5~{\rm M_\odot}$, $63.10~{\rm d})$ and $(1.4~{\rm M_\odot}$, $1.5~{\rm M_\odot}$, $0.79~{\rm d})$, we find the final WD masses to be $\sim 0.27~{\rm M_\odot}$ and $\sim 0.17~{\rm M_\odot}$ in Model~\textit{TEW5}, and $\sim 0.4~{\rm M_\odot}$ and $\sim 0.24~{\rm M_\odot}$ in Model~\textit{NoWind}, respectively.  

In \autoref{fig:MPorb}, we show the ten evolutionary tracks in the donor mass--orbital period ($M_{\rm d}$--$P_{\rm orb}$) plane. The red solid line in each panel shows the $M_{\rm WD}$--$P_{\rm orb}$ relation calculated by \cite{Lin+2011}. We see that the final state of the tracks are on the left of the red solid line, that is, the final HeWD get less massive at given orbital period with increasing wind mass loss.

\begin{figure*}
    \centering
    \includegraphics[width=\linewidth]{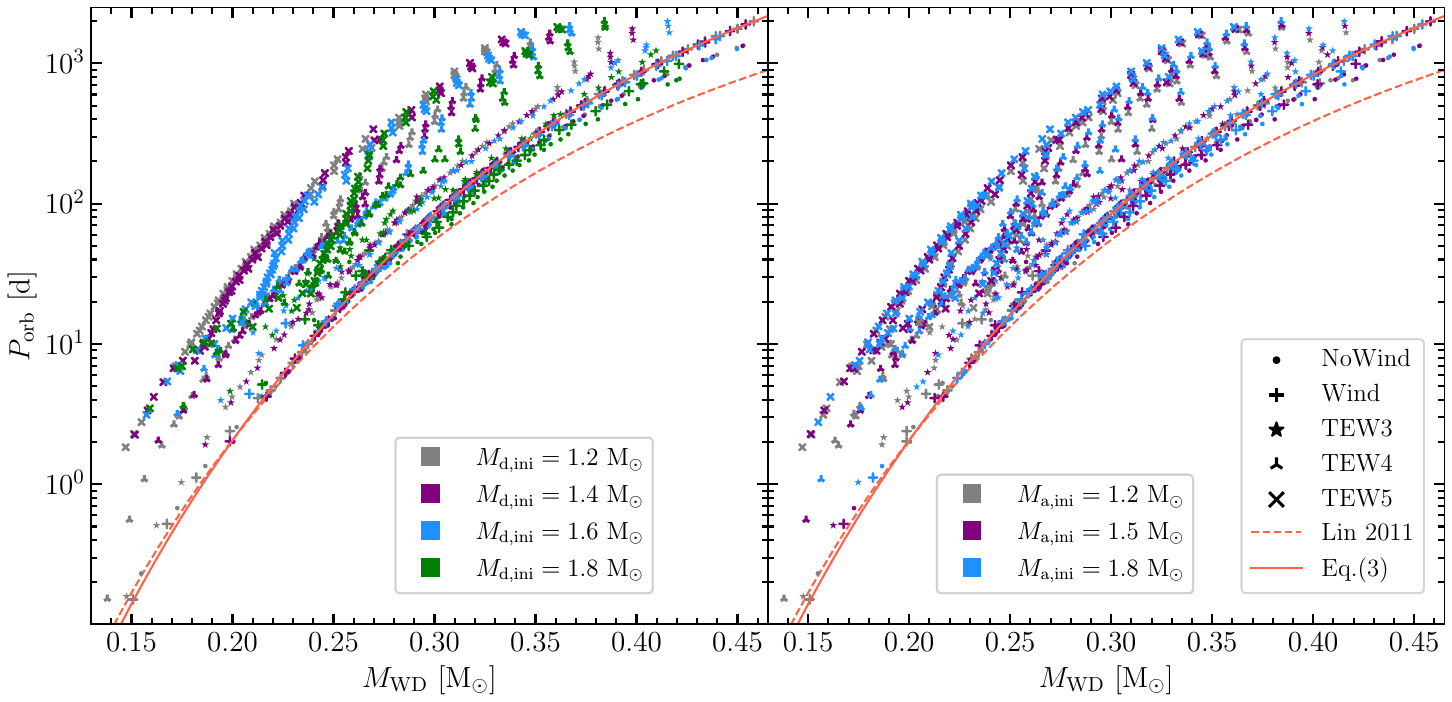}
    \caption{The $M_{\rm WD}$--$P_{\rm orb}$ relations {for the five sets of binary models}. The dots, pluses, stars, triadius and crosses correspond to Models \textit{NoWind, Wind, TEW3, TEW4 {\rm and} TEW5}, respectively. The red solid and dashed line indicate the best-fitting $M_{\rm wd}$--$P_{\rm orb}$ relations of Model~\textit{Wind} (\autoref{eq:MwdPorbRel}) and of \protect\cite{Lin+2011}, respectively. In the left panel, the gray, purple, blue and green markers correspond to $M_{\rm d,ini}=1.2,~1.4,~1.6$ and $1.8~{\rm M_\odot}$, respectively. {In the right panel, the gray, purple and blue markers correspond to $M_{\rm a,ini}=1.2,~1.5$ and $1.8~{\rm M_\odot}$, respectively.} \label{fig:Relation}}
\end{figure*}

The $M_{\rm WD}$--$P_{\rm orb}$ relations of different models are compared in the left panel of \autoref{fig:Relation}. The dots, pluses, stars, triadius and crosses represent the results in Models \textit{NoWind, Wind, TEW3, TEW4 {\rm and} TEW5}, respectively. In the left panel, the gray, purple, blue and green markers correspond to $M_{\rm d,ini}=1.2,~1.4,~1.6$ and $1.8~{\rm M_\odot}$, respectively. { In the right panel, the gray, purple and blue markers correspond to $M_{\rm a,ini}=1.2,~1.5$ and $1.8~{\rm M_\odot}$, respectively. Because of the weak dependence of the Roche-lobe radius on the mass ratio in our cases, the $M_{\rm WD}$--$P_{\rm orb}$ relation is also weakly dependent on the initial accretor mass.} The relation obtained by \cite{Lin+2011} is shown by the red dashed line. The difference between our results in Model~\textit{NoWind} and \cite{Lin+2011} is caused by the different RLOF schemes used \citep[see discussion in][]{Zhang+2021}. The red solid line in \autoref{fig:Relation} denotes the best-fit curve for Model~\textit{Wind} in the form\footnote{The fitting was made by using the least squares method implemented in \textsc{SciPy} \citep{Virtanen+2020}.}
\begin{equation}\label{eq:MwdPorbRel}
    P_{\rm orb}({\rm d})=\frac{249831.9489m^{9.3420}}{(0.00128+1.5048m^{7.9567}+1.4975m^{7.9577})^{0.5}},
\end{equation}
where $m=M_{\rm WD}/{\rm M_\odot}$. Note that there is a break between binaries with $M_{\rm WD}\lesssim 0.21~{\rm M_\odot}$ and $\gtrsim 0.21~{\rm M_\odot}$ in Models~\textit{NoWind} and \textit{Wind}, because the progenitors of HeWDs with mass $\lesssim 0.21~{\rm M_\odot}$ do not have a well-developed degenerate helium core at the onset of mass transfer \citep[e.g.,][]{Ergma+1998,Nelson+2004,Jia+2014,Istrate+2016,ChenXF+2017}. Although our fitting is limited for HeWDs with $M_{\rm WD}>0.21~{\rm M_\odot}$, it also presents a good description of the $M_{\rm WD}$--$P_{\rm orb}$ relation for $M_{\rm WD}<0.21~{\rm M_\odot}$.

For Models~\textit{TEW3}, \textit{TEW4} and \textit{TEW5}, $M_{\rm WD}$ and $P_{\rm orb}$ do not follow a simple relation as for Models~\textit{NoWind {\rm and} Wind}. The deviation becomes larger with increasing wind mass loss and the $M_{\rm WD}$--$P_{\rm orb}$ relation moves to up left of the red line. In these cases the donors do not always fill their RLs, so the descended HeWD mass depend on both the initial orbital period and the donor mass -- $M_{\rm WD}$ increases with increasing $M_{\rm d, ini}$. The final orbital periods in Model~\textit{TEW5} are about {one} order of magnitude longer than those in Model~\textit{NoWind} for a given $M_{\rm WD}$, and there is no HeWD binary formed with $P_{\rm orb}\lesssim 2~{\rm d}$ in Model~\textit{TEW5}.

\section{Comparison with observations}\label{sec:com}

\begin{table*}
    \centering
    \setlength{\tabcolsep}{5pt}
    \caption{Masses ($M_{\rm WD}$), orbital periods ($P_{\rm orb}$), eccentricities ($e$), NS masses ($M_{\rm NS}$) and NS spin periods ($P_{\rm s}$) for HeWD+NS binaries. Note that J0337+1715 is a hierarchical triple system. The binaries are sorted by the magnitude of the orbital periods.\label{tab:obspsr}}
    \begin{tabular}{llllllll}
         \hline
         Name
         &Type
         &$M_{\rm WD}~{\rm [M_\odot]}$
         &$P_{\rm orb}~{\rm [d]}$
         &$e$
         &$M_{\rm NS}~{\rm [M_\odot]}$
         &$P_{\rm s}~{\rm [ms]}$
         &References\\
         \hline
    J2016+1948&HeWD+NS&$0.45\pm0.02$&$635.04$&$0.00147981$&$1.0\pm0.5$&$64.9404$&\cite{Gonzalez+2011}\\
    J0337+1715b&HeWD+(HeWD+NS)&$0.41058\pm0.00040$&$327.25539$&$0.035290$&$\sim 1.64$&-&\cite{Ransom+2014}\\
    J1713+0747&HeWD+NS&$0.292\pm0.011$&$67.82513$&$7.494\times10^{-5}$&$1.35\pm0.07$&$4.57$&\cite{Arzoumanian+2018}\\
    J1910+1256&HeWD+NS&$0.315\pm0.015$&$58.47$&$2.302\times10^{-4}$&$1.6\pm0.6$&$4.98358$&\cite{Gonzalez+2011}\\
    J0614--3329&HeWD+NS&$0.24\pm0.04$&$53.585$&-&-&$3.14867$&\cite{Bassa+2016}\\
    J2234+0611&HeWD+NS&$0.298_{-0.012}^{+0.015}$&$32.00140$&$0.129274$&$1.353_{-0.017}^{+0.014}$&$3.57658$&\cite{Stovall+2019}\\
    J1946+3417&HeWD+NS&$0.2656\pm0.0019$&$27.01995$&$0.134495$&$1.828\pm0.022$&$3.17014$&\cite{Barr+2017}\\
    J0955--6150 &HeWD+NS&$0.254\pm0.002$&$24.5784$&$0.11750575$&$1.71\pm0.02$&$1.99936$&\cite{Serylak+2022}\\
    J1950+2414&HeWD+NS&$0.2788\pm{0.0038}$&$22.191371$&$0.07981173$&$1.500\pm0.022$&$4.30478$&\cite{Zhu+2019}\\
    B1855+09&HeWD+NS&$0.244_{-0.012}^{+0.014}$&$12.32717$&$2.170\times10^{-5}$&$1.37_{-0.10}^{+0.13}$&$5.36$&\cite{Arzoumanian+2018}\\
    J1918--0642 &HeWD+NS&$0.231\pm0.010$&$10.91318$&$2.03\times10^{-5}$&$1.29_{-0.09}^{+0.10}$&$7.65$&\cite{Arzoumanian+2018}\\
    J1811--2405&HeWD+NS&$0.31_{-0.06}^{+0.08}$&$6.272302$&$1.11\times10^{-6}$&$2.0_{-0.5}^{+0.8}$&$2.66059$&\cite{Ng+2020}\\
    J0437--4715&HeWD+NS&$0.224\pm0.007$&$5.7410459$&$1.918\times10^{-5}$&$1.44\pm0.07$&$5.75745$&\cite{Reardon+2016}\\
    J0740+6620&HeWD+NS&$0.253_{-0.005}^{+0.006}$&$4.76694$&$6\times10^{-6}$&$2.08\pm0.07$&$2.88574$&\cite{Fonseca+2021}\\
    J0337+1715a&HeWD+NS&$0.19780\pm0.00019$&$1.62940$&$0.00069890$&$1.4401\pm0.0015$&$2.73259$&\cite{Ransom+2014}\\
    J1909--3744&HeWD+NS&$0.205\pm0.003$&$1.53345$&$1.15\times10^{-7}$&$1.45\pm0.03$&$2.94711$&\cite{Shamohammadi+2023}\\
    J2043+1711&HeWD+NS&$0.173\pm0.010$&$1.48229$&-&$1.38_{-0.13}^{+0.12}$&$2.38$&\cite{Arzoumanian+2018}\\
    J1910--5958A&HeWD+NS&$0.202\pm0.006$&$0.837113$&-&$1.55\pm 0.07$&$3.26619$&\cite{Corongiu+2023}\\
    J1012+5307&HeWD+NS&$0.165\pm0.015$&$0.60467$&$0.0000012$&$1.72\pm16$&$5.25575$&\cite{Sanchez+2020}\\
    J1738+0333&HeWD+NS&$0.181_{-0.005}^{+0.007}$&$0.35479$&$3.5\times10^{-7}$&$1.47_{-0.06}^{+0.07}$&$5.85 $&\cite{Antoniadis+2012}\\
    J0751+1807&HeWD+NS&$0.16\pm0.01$&$0.26314$&$0.0000005$&$1.64\pm0.15$&$3.47877$&\cite{Desvignes+2016}\\
    J0348+0432&HeWD+NS&$0.172\pm0.003$&$0.10242$&$0.0000020$&$2.01\pm0.04$&$39.1227$&\cite{Antoniadis+2013}\\
    \hline
    \end{tabular}
\end{table*}

\begin{figure*}
    \centering
    \includegraphics[width=\linewidth]{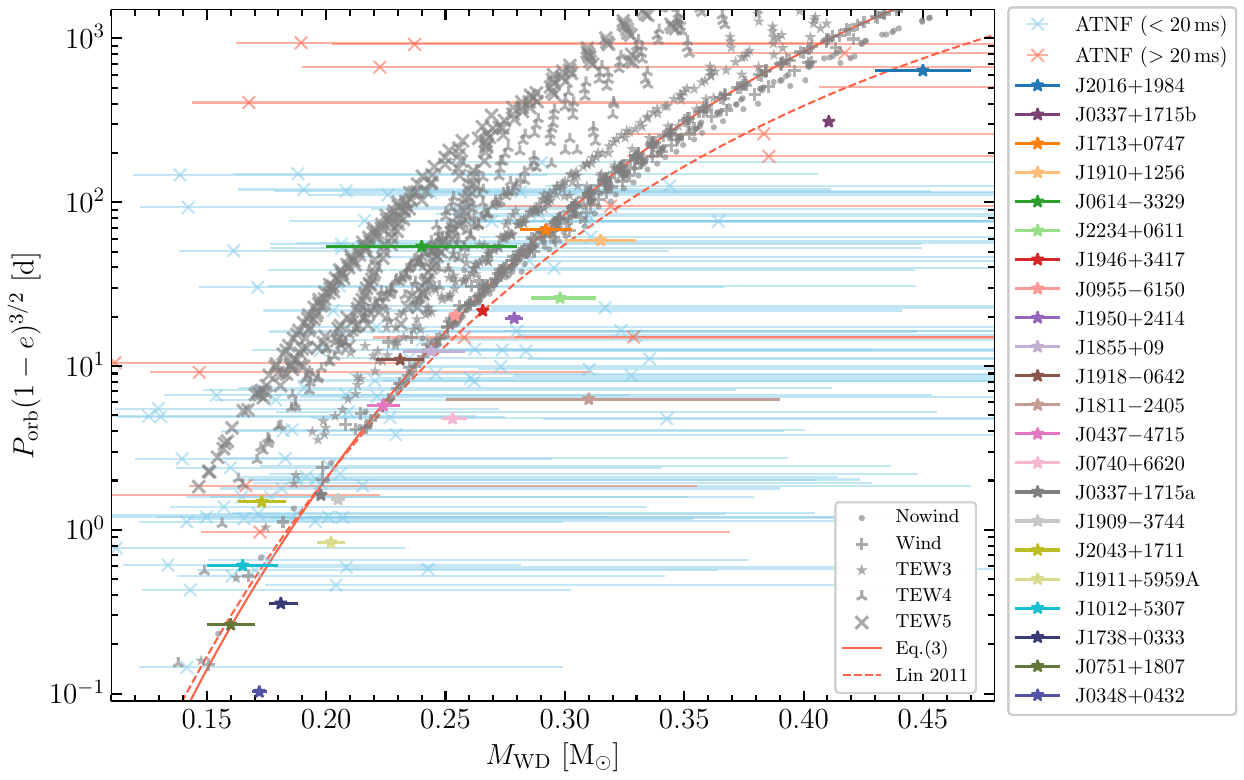}
    \caption{{Comparison between observational data and the results obtained from our models in the $M_{\rm WD}$--$P_{\rm orb}$ plane.} Similar to \autoref{fig:Relation}, the gray markers show the $M_{\rm WD}$--$P_{\rm orb}$ relations of our models. The light {blue} and the light {red} crosses indicate the observed binary pulsars with spin periods $<20~{\rm ms}$ and $>20~{\rm ms}$, respectively. The colored stars show the HeWDs in binary pulsars with more accurate mass measurements (see \autoref{tab:obspsr}).\label{fig:psr}}
\end{figure*}

\begin{figure}
    \centering
    \includegraphics[width=\linewidth]{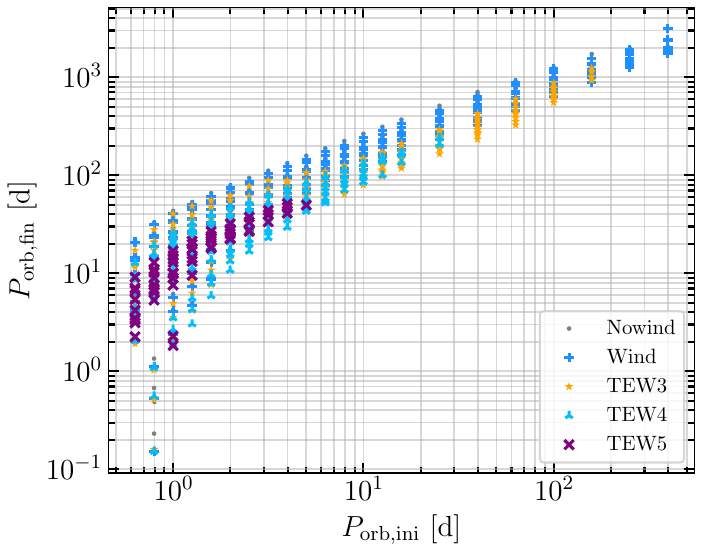}
    \caption{The relation between the initial orbital period $P_{\rm orb,ini}$ and the final period $P_{\rm orb,fin}$ in different models with the NS accreted mass via RLOF $\geq 0.1 \rm M_\odot$.}
    \label{fig:psrpipf}
\end{figure}

Most HeWD+NS binaries are binary MSPs with spin periods $P_{\rm s}\lesssim 20~{\rm ms}$. The short spin periods are thought to origin from accretion of mass and angular momentum of the NSs during their previous low/intermediate-mass X-ray binary evolution \citep[][for a review]{Bhattacharya+1991}.

We collect the data of HeWD+NS binaries from the ATNF pulsar catalogue \citep{Manchester+2005}\footnote{\url{https://www.atnf.csiro.au/people/pulsar/psrcat/}}. We plot the measured masses and orbital periods of HeWD+NS binaries in \autoref{fig:psr}. They generally have a small but non-zero eccentricity, so the observed orbital periods should be reduced by a factor of $(1-e)^{3/2}$ (where $e$ is the eccentricity) when compared with the calculated $M_{\rm WD}$--$P_{\rm orb}$ relation. The blue crosses represent the binaries with the WD masses calculated by assuming the NS masses to be $1.35~{\rm M_\odot}$ and the orbital inclination angles to be $60\degree$ (the light {blue} and light {red} ones correspond to the spin periods $<20~{\rm ms}$ and $>20~{\rm ms}$, respectively). The lower and upper limits of the WD masses are obtained by assuming the orbital inclination angle to be $90\degree$  and $26\degree$ respectively, which cover the $90\%$ probability mass range for randomly oriented orbits. The coloured stars indicate the binary pulsars with more accurate mass measurements, including by measuring the orbital decay caused by gravitational wave radiation \citep{Antoniadis+2013} and by measuring the Shapiro delay in pulsar timing \citep{Ng+2020}. The parameters and references of these binaries are listed in \autoref{tab:obspsr}.

From \autoref{fig:psr}, we find that quite a few HeWD+NS binaries are located on the left of the traditional $M_{\rm WD}$--$P_{\rm orb}$ relation depicted by the red dashed line, that is, the observed WD masses are less than expected at given orbital period, although the error bars of the WD masses are quite large. 
{Note that small values of the orbital inclination and
large NS mass can increase the inferred HeWD mass for given observed mass
functions. However, \cite{1996A&A...315..453T} pointed out that there does not seem to be any observational
selection effect favouring small inclination angle for HeWD+NS binaries. \cite{Shao+2012} accordingly} suggested that, to explain the {mismatch} between the observations with the theoretical $M_{\rm WD}$--$P_{\rm orb}$ relation, an initially massive NS ($\sim 2~{\rm M_\odot}$) may be required for part of the binaries\footnote{
The reason is that accretion is likely inefficient in wide binaries due to the super-Eddington mass transfer and the thermal and viscous instability in the accretion disc \citep{Lasota+2001,Hameury+2020R}, thus the NSs have grown little mass during the mass transfer stage.}.
Considering the fact that the mass measurements of the recycled pulsars reveal a mean mass of $1.48~{\rm M_\odot}$ and a dispersion of $0.2~{\rm M_\odot}$ \citep{Ozel+2012}, it is unclear whether and why the birth masses of NSs in specific binaries are more massive than others.

\autoref{fig:psr} shows that models with wind mass loss seem to offer a more reasonable explanation for the {least massive} HeWD binaries \citep[see also][]{2017MNRAS.464..237S}. For example, the masses of HeWDs in PSRs~J1713+0747, J0955--6150 and J1918--0642 are significantly lower than expected from the red dashed curve calculated by both \cite{Lin+2011} and Model~\textit{Nowind}, but consistent with Model~\textit{Wind}. Moreover, PSRs~J0614--3329 and J2043+1711 can only be accounted for by the TEW models, suggesting that {it is necessary to consider the effect of wind mass loss on the $M_{\rm WD}$--$P_{\rm orb}$ relation}\footnote{{There are a few HeWD binaries located to the right of the red solid/dashed curves. The HeWDs in these binaries could be formed from lower-metallicity stars. Stars with the same initial mass but lower
metallicities have smaller stellar radius and form WDs with shorter orbital periods \citep{Rappaport+1995,Tauris+1999,Jia+2014,Istrate+2016}.}}. 

An issue related to the TEW models is whether the NS spins can be accelerated to milliseconds with wind accretion. 
\autoref{fig:RRL} shows that in Models~\textit{TEW4} and \textit{TEW5}, the RGB stars do not always fill their RLs, thus the NSs can only accrete mass and angular momentum from the stellar wind in this situation. The relatively low accretion efficiency in wind-fed X-ray binaries makes it hard for the NSs to be efficiently spun up. \autoref{fig:psrpipf} {shows} the initial orbital period vs. the final orbital period relation in various models. Only the binaries with RLOF accreted mass $\geq 0.1~{\rm M_\odot}$ \citep[which is necessary for the formation of MSPs under disc accretion,][]{Burderi+2005} are plotted. It is evident that, the more wind mass loss, the shorter the final orbital periods of MSPs. {In 
Models~\textit{TEW4} and \textit{TEW5}, the longest orbital periods for MSPs are around $200~{\rm d}$ and $50~{\rm d}$, respectively. They seem to be in contradiction with the observed MSPs demonstrated 
in \autoref{fig:psr} and \autoref{tab:obspsr}, especially for Model~\textit{TEW5}.}

A potential solution to this problem is the wind-RLOF scheme, where the slowly expanding stellar wind fills the RL and is gravitationally focused to the $L_{\rm 1}$ point \citep{Mohamed+2012,Abate+2013}. Compared with the canonical Bondi-Hoyle-Lyttleton wind accretion \citep{Hoyle+1939,Bondi+1944}, wind-RLOF may significantly enhance the mass and angular momentum transfer efficiencies, and possibly spin up the NSs to be MSPs. The condition and mechanisms of wind-RLOF depend on the properties of dust grains in the cool, outer atmosphere of AGB stars. When the dust grains are accelerated by radiation pressure, collisional momentum transfer between dust and gas also drags the surrounding gas outward.
\cite{Mohamed+2012} and \cite{Abate+2013} suggested that wind-RLOF occurs when the dust-formation radius is beyond the RL radius of the AGB star. It is still uncertain whether RGB stars also have wind dust-formation regions \citep{Boyer+2008,Boyer+2010,Origlia+2010,McDonald+2011a,McDonald+2011b}. Whether and how wind-RLOF can alter the spin evolution of NSs need further investigation and are beyond the scope of this work.

\begin{figure*}
    \centering
    \includegraphics[width=\linewidth]{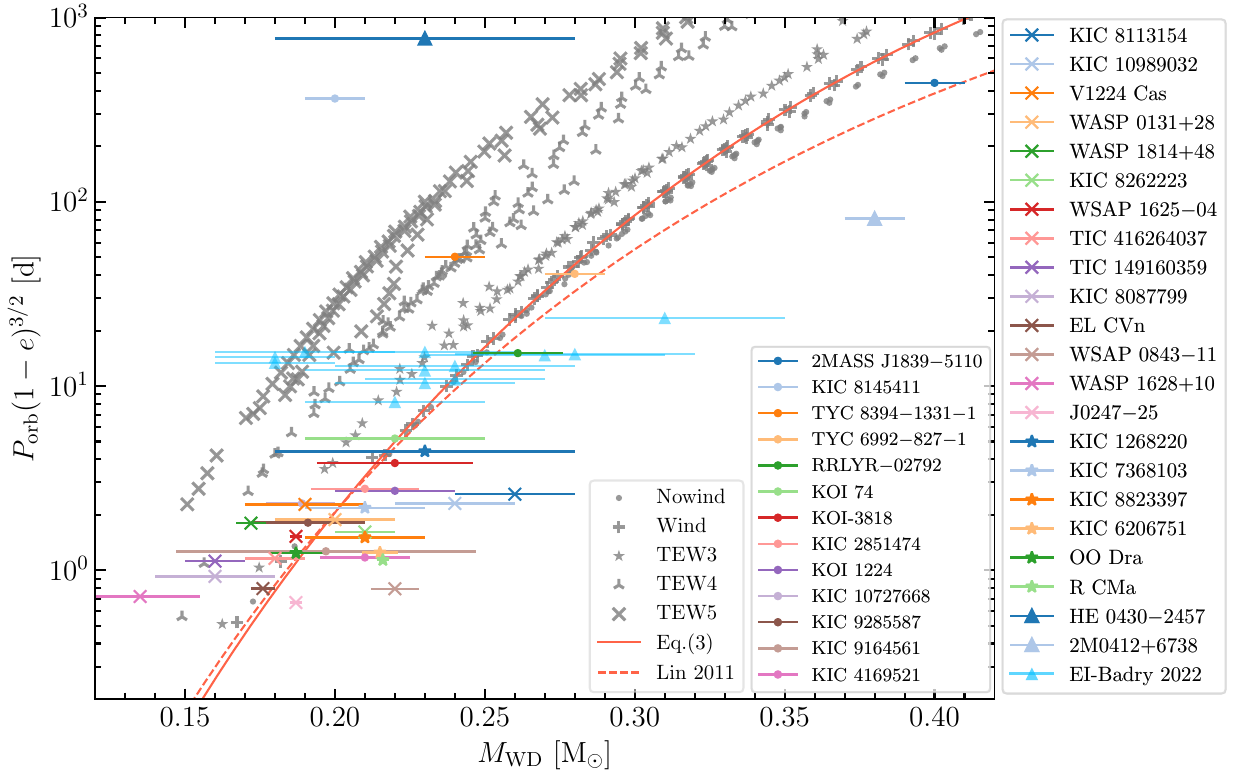}
    \caption{Similar to \autoref{fig:psr} but for HeWD+MS binaries. {Only binaries with $M_{\rm a,ini}<M_{\rm d,ini}$ are shown.} The dots, crosses, stars and triangles represent the HeWD binaries, EL~CVn-type binaries, R~CMa-tpye binaries and other preWD+MS/subgiant binaries, respectively.
    \label{fig:ms}}
\end{figure*}

\begin{table*}
    \centering
    \setlength{\tabcolsep}{6pt}
    \caption{Similar to \autoref{tab:obspsr} but for (pre)HeWD+MS/subgiant binaries. Here, $M_{\rm MS}$ represent the MS/subgiant mass. These binaries are sorted by the magnitude of the orbital periods for each subclass. \label{tab:obsMS}}
    \begin{tabular}{lllllll}
         \hline
         Name
         &Type
         &$M_{\rm WD}~{\rm [M_\odot]}$
         &$P_{\rm orb}~{\rm [d]}$
         &$e$
         &$M_{\rm MS}~{\rm [M_\odot]}$
         &References\\
\hline

    2MASS~J1836--5110&HeWD+subgiant&$0.40\pm0.01$&$461.48\pm0.04$&$0.028\pm0.001$&$1.38\pm0.16$&\cite{Parsons+2022}\\ KIC~8145411&HeWD+MS&$0.20\pm0.01$&$455.83$&$0.143$&$1.132\pm0.08$&\cite{Masuda+2019}\\
    TYC~8394--1331--1 &HeWD+subgiant&$0.24\pm0.01$&$51.851$&$0.02\pm0.02$&$1.31\pm0.12$&\cite{Parsons+2022}\\
    TYC~6992--827--1&HeWD+subgiant&$0.28\pm0.01$&$41.45\pm0.01$&$0.013\pm0.06$&$1.31\pm0.14$&\cite{Parsons+2022}\\
    RRLYR-02792&HeWD+subgiant&$0.261\pm0.015$&$15.24340$&$0.0072$&$1.67\pm0.06$&\cite{Pietrzynski+2012}\\
    KOI~74&HeWD+MS&$0.22\pm0.03$&$5.18875$&-&$2.22_{-0.14}^{+0.10}$&\cite{vanKerkwijk+2010}\\
    KOI--3818&HeWD+MS&$0.220\pm0.026$&$3.8170428$&-&$2.14\pm0.12$&\cite{Faigler+2015}\\
    KIC~2851474&HeWD+MS&$0.210\pm0.018$&$2.7682925$&-&$2.34\pm0.19$&\cite{Faigler+2015}\\
    KOI~1224&HeWD+MS&$0.22\pm0.02$&$2.69802$&-&$1.59\pm0.06$&\cite{Breton+2012}\\
    KIC~10727668&HeWD+MS&$0.189^{+0.011}_{-0.012}$&$2.30592$&$0.018_{-0.007}^{+0.002}$&$3.03^{+0.12}_{-0.17}$&\cite{Odesse+2022}\\
    KIC~9285587&HeWD+MS&$0.191\pm0.019$&$1.8119579$&-&$1.94\pm0.16$&\cite{Faigler+2015}\\
    KIC~9164561&HeWD+MS&$0.197\pm0.05$&$1.267040$&-&$2.02\pm0.06$&\cite{Rappaport+2015}\\KIC~4169521&HeWD+MS&$0.210\pm0.015$&$1.172555671$&-&$1.982\pm0.092$&\cite{Faigler+2015}\\
\hline
    KIC~8113154&EL~CVn-type&$0.26\pm0.02$&$2.5868779$&-&$1.65\pm0.01$&\cite{Zhang+2019}\\
    KIC~10989032&EL~CVn-type&$0.24\pm0.02$&$2.3050976$&-&$2.64\pm0.19$&\cite{zhang+2017}\\
    V1224~Cas&EL~CVn-type&$0.19\pm0.02$&$2.27537$&-&$2.16\pm0.02$&\cite{Wangkun+2018}\\
    WASP~0131+28&EL~CVn-type&$0.200\pm0.02$&$1.88275$&-&$1.986\pm0.017$&\cite{Lee+2020}\\
    WASP~1814+48&EL~CVn-type&$0.172\pm0.005$&$1.79942$&-&$1.659\pm0.048$&\cite{Lee+2022}\\
    KIC~8262223&EL~CVn-type&$0.21\pm0.01$&$1.61301476$&-&$1.96\pm0.06$&\cite{Guo+2017}\\
    TIC~416264037&EL~CVn-type&$0.18\pm0.01$&$1.15991$&-&$1.70\pm0.10$&\cite{Wang+2020}\\
    WASP~1625--04&EL~CVn-type&$0.187\pm0.002$&$1.526323$&-&$1.745\pm0.013$&\cite{Lee+2022c}\\
    TIC~149160359&EL~CVn-type&$0.16\pm0.01$&$1.120738$&-&$1.75\pm0.10$&\cite{Wang+2020}\\
    KIC~8087799&EL~CVn-type&$0.16\pm0.02$&$0.9262976$&-&$2.20\pm0.20$ &\cite{zhang+2017}\\
    EL~CVn&EL~CVn-type&$0.176\pm0.004$&$0.795627$&-&$1.430\pm0.027$&\cite{Wangluqian+2020}\\
    WASP~0843--11&EL~CVn-type&$0.220\pm0.008$&$0.792839$&-&$1.733\pm0.031$&\cite{Hong+2021}\\
    WASP~1628+10&EL~CVn-type&$0.135\pm0.02$&$0.72$&-&$1.36\pm0.05$&\cite{Maxted+2014}\\
    J0247--25&EL~CVn-type&$0.187\pm0.002$&$0.6678306$&-&$1.356\pm0.007$&\cite{Kim+2021}\\
\hline
    KIC~12268220&R~CMa-type&$0.23\pm0.05$&$4.421580$&-&$1.99_{-0.40}^{+0.52}$&\cite{Cui+2020}\\
    KIC~7368103&R~CMa-type&$0.21\pm0.02$&$2.1825147$&-&$1.77\pm0.19$&\cite{Wang+2019}\\
    KIC~8823397&R~CMa-type&$0.21\pm0.02$&$1.5065038$&-&$2.09\pm0.17$&\cite{Wang+2019}\\
    KIC~6206751&R~CMa-type&$0.215\pm0.006$&$1.24534$&-&$1.66\pm0.04$&\cite{Lee+2018b}\\
    OO~Dra&R~CMa-type&$0.187\pm0.009$&$1.23838$&-&$2.031\pm0.058$&\cite{Lee+2018}\\
    R CMa&R~CMa-type&$0.216\pm0.002$&$1.135956$&-&$1.67\pm0.01$&\cite{Lehmann+2018}\\
\hline
    HE~0430--2457&preHeWD+MS&$0.23\pm0.05$&$771\pm3$&-&$0.71\pm0.09$&\cite{Vos+2018}\\
    2M04123153+6738486&SRG+subgiant&$0.38\pm0.01$&$81.2$&-&$1.91\pm0.03$&\cite{Jayasinghe+2022}\\
    448452383082046208&SRG+MS&$0.31\pm0.04$&$23.436$&-&$2.4\pm0.2$&\cite{El-Badry+2022b}\\
    6000420920026118656&SRG+MS&$0.19\pm0.03$&$15.312$&-&$2.2\pm0.2$&\cite{El-Badry+2022b}\\
    1850548988047789696&SRG+MS&$0.23\pm0.03$&$15.302$&-&$2.3\pm0.1$&\cite{El-Badry+2022b}\\
    5243109471519822720&SRG+MS&$0.28\pm0.04$&$14.914$&-&$2.0\pm0.2$&\cite{El-Badry+2022b}\\
    2933630927108779776&SRG+MS&$0.27\pm0.04$&$14.715$&-&$2.2\pm0.2$&\cite{El-Badry+2022b}\\
    6734611563148165632&SRG+MS&$0.18\pm0.02$&$14.344$&-&$2.2\pm0.1$&\cite{El-Badry+2022b}\\
    548272473920331136&SRG+MS&$0.18\pm0.02$&$13.379$&-&$1.9\pm0.1$&\cite{El-Badry+2022b}\\
    5694373091078326784&SRG+MS&$0.24\pm0.04$&$12.876$&-&$2.1\pm0.2$&\cite{El-Badry+2022b}\\
    5536105058044762240&SRG+MS&$0.23\pm0.04$&$12.177$&-&$1.9\pm0.2$&\cite{El-Badry+2022b}\\
    2219809419798508544&SRG+MS&$0.24\pm0.03$&$10.865$&-&$2.0\pm0.1$&\cite{El-Badry+2022b}\\
    2966694650501747328&SRG+MS&$0.23\pm0.03$&$10.398$&-&$2.0\pm0.1$&\cite{El-Badry+2022b}\\
    948585824160038912&SRG+MS&$0.22\pm0.03$&$8.202$&-&$2.0\pm0.1$&\cite{El-Badry+2022b}\\
\hline
    \end{tabular}
\end{table*}

Another possible mechanism that can affect the $M_{\rm WD}$--$P_{\rm orb}$ relation is evaporation of the progenitor of a HeWD by high-energy irradiation from an MSP \citep{vandenHeuvel+1988,Ruderman+1998}. \cite{Jia+2016} showed that this can make the $M_{\rm WD}$--$P_{\rm orb}$ relation more scattered, in particular in binaries with $P_{\rm orb}\lesssim20~{\rm d}$. So, it is difficult to account for the HeWD binaries like PSRs J0614--3329 ($P_{\rm orb}=53.6$ d) and J1713+0747 ($P_{\rm orb}=67.8$ d) {by donor evaporation}. {Furthermore}, some HeWD+MS binaries, in which high-energy irradiation is obviously lacking, also seem to deviate from the $M_{\rm WD}$--$P_{\rm orb}$ relation. To see this in more detail, we compare the data of various types of HeWD binaries with theoretical predictions in \autoref{fig:ms}. {Here, we only show binaries with $M_{\rm a,ini}<M_{\rm d,ini}$ to ensure that the MS donors (the progenitors of HeWDs) evolve faster than the MS accretors.} {We} collect the data of HeWD+MS binaries from the literature and list them in \autoref{tab:obsMS}. Apart from the canonical HeWD+MS binaries we also include EL~CVn-type binaries, R~CMa-type binaries and stripped red giant (SRG)+MS binaries\footnote{Here, EL~CVn-type binaries are eclipsing binaries containing an A/F-type MS dwarf star and a low-mass ($\lesssim0.2~{\rm M_\odot}$) HeWD precursor (preHeWD) \citep{Maxted+2014}. They are thought to form from stable mass transfer rather than common envelope evolution, because the later usually leads to merging of the components if the RGB stars have such low mass cores \citep{ChenXF+2017}. R~CMa-type binaries are the progenitors of EL~CVn binaries, and have a short orbital period and a low-mass ratio among Algol systems \citep{Varricatt+1999,Budding+2011,Lehmann+2013,Lee+2016}. SRG+MS binaries in \autoref{tab:obsMS}, selected from \textit{Gaia}~DR3 \citep{El-Badry+2022b}, are earlier than R~CMa-type binaries and will also evolve to be HeWD+MS binaries. The donors are close to filling their RLs.}. In addition, three HeWD+subgiant binaries and one SRG+subgiant binary are also listed in \autoref{tab:obsMS}. Although the subgiants have evolved away from MS, the mass transfer process has not occurred. So, they are also expected to follow  the the $M_{\rm WD}$--$P_{\rm orb}$ relation. 

\autoref{fig:ms} shows that, while the majority of the sample seem to be compatible with the predictions of Models~\textit{NoWind} and~\textit{Wind} (corresponding to the red dashed and solid lines respectively), there are quite a few outliers located on the left of the lines, which are more consistent with the TEW models. For example, to account for TYC~8394--1331--1, WASP~1814+6738, TIC~149160359, KIC~8087799, WASP~1628+10, and several anomalous SRG+HeWD binaries requires models with $B_{\rm W}=10^3-10^4$. Moreover, HE~0430--2457 and KIC~8145411 have very long orbital periods (771 d and 456 d, respectively). The HeWD masses are so small that even Model~\textit{TEW5} cannot explain them. These two peculiar binaries may have formed from triple/multiple interactions \citep[][]{Vos+2018,Masuda+2019,Gosnell+2019,Pandey+2021}. However, this scenario can not explain why both KIC~8145411 and HE 0430–2457 have nearly circular orbits\footnote{The probability of falsely rejecting a circular orbit is $\mathcal{P} = 0.59$ for HE~0430–2457 \cite{Vos+2018}. \cite{Lucy+1971} argued that if $\mathcal{P}>0.05$, the orbit is effectively circular.} and why low-mass HeWDs do not appear to be very rare among long-period binaries \citep[see][]{Masuda+2019}. In addition, \cite{Khurana+2022} suggested that KIC~8145411 may be formed by binary-binary strong encounters in star clusters, where the binary orbit was significantly widened. But they also pointed out that the formation rate in this channel is rather low, only $\sim 4~{\rm Myr^{-1}}$.

\section{Discussion and Conclusions}\label{sec:dis}

In this work, we investigate the effects of wind mass loss on the $M_{\rm WD}$--$P_{\rm orb}$ relation. We show that enhanced stellar winds significantly influence the formation paths of HeWDs, because wind mass loss continues stripping the envelope of the RGB progenitors of HeWDs or the pre-HeWDs during and after the RLOF phase. These processes can efficiently slow the growth of the degenerate helium core of the RGB stars. The TEW model may explain why some MSP binaries and HeWD+MS binaries possess less massive HeWD companions than expected. 

\cite{Tout+1988} proposed the idea of TEW on a heuristic basis to account for the RS~CVn system Z~Her, which was the only reliable system {where} a mass-radius paradox certainly existed. Afterwards, this prescription has been widely invoked to interpret the formation and evolution of Algol-like and related binaries \citep{Tout+1988b,Hanna+2012,Zhang+2013}, Barium stars \citep{Han+1995,Jorissen+1998,Jorissen+2000,Karakas+2000,Bonacic+2008,Lv+2020} and Carbon-rich extremely metal poor stars \citep{Lau+2007}, the morphology of the off-centre planetary nebulae \citep{Soker+1998}, Type Ia supernovae from symbiotic stars \citep{Scott+1993,Chen+2011,Meng+2016,Wu+2021}, the horizontal branch morphology of globular clusters \citep{Lei+2013,Lei+2013b,Lei+2014}, and long-period eccentric binaries with a barium star \citep{Bonacic+2008,Vos+2015}, HeWD \citep{Siess+2014,Merle+2014} or subdwarf star companion \citep{Vos+2015}. However, the mechanism of TEW is still under debate. Apart from tidal interactions or dynamo activity as suggested by \cite{Tout+1988}, stellar winds of RGB/AGB stars in  binary systems might also be enhanced by stellar pulsation \citep{Eggleton+2002}, rotation \citep{Nieuwenhuijzen+1988,Zamanov+2008,Bear+2010} and the effective gravity reduced by the presence of a companion star \citep{Frankowski+2001}. Thus, it is premature to confidently estimate the magnitude of $B_{\rm w}$ (see \autoref{eq:reimers} and \ref{eq:tew}). It likely varies across different types of binaries and different evolutionary stages.  \autoref{fig:psr} and \ref{fig:ms} demonstrate that the sample with parameters accurate enough to test the $M_{\rm WD}$--$P_{\rm orb}$ relation is still small. There are not yet enough HeWD binaries to make a more statistically significant judgement and to find that under which condition can the TEW model work.

In addition, since mass loss via TEW proceeds with RLOF, this may result in an eccentric orbit. \cite{Soker+2000} proposed that the significant eccentricities $\sim 0.1-0.4$ found in some tidally strongly interacting binaries (e.g., binaries with RGB/AGB companions) are caused by an enhanced mass loss during periastron passages. \cite{Bonacic+2008} found that the enhanced mass loss of the AGB stars at periastron can result in a net eccentricity growth rate that is comparable to the tidal circularisation. Furthermore, \cite{Siess+2014} showed that an enhanced-wind mass loss on the RGB can avoid RLOF and they found that the eccentricity can be preserved and even increased if the initial separation is large enough. Although {here} are HeWD binaries like KIC~8145411 {with} relatively large eccentricities, most wide binaries in Tables~\ref{tab:obspsr} and \ref{tab:obsMS} are almost circular. Future observations of more wide HeWD binaries are crucial in testing the applicability of the TEW model.

\section*{Acknowledgements}
We are grateful to an anonymous referee for helpful comments. This work was supported by the National Key Research and Development Program of China (2021YFA0718500), the Natural Science Foundation of China under grant No. 12041301, 12121003, and Project U1838201 supported by NSFC and CAS. The computation was made by using the facilities at the High-Performance Computing Center of Collaborative Innovation Center of Advanced Microstructures (Nanjing University, \url{https://hpc.nju.edu.cn}). Figures in this work are made use of \textsc{Matplotlib} \citep{matplotlib}, \textsc{NumPy} \citep{numpy} and \textsc{Jupyter-Lab} (\url{https://jupyter.org}).

\section*{Data Availability}
All data underlying this article will be shared on reasonable request to the corresponding author.

\bibliographystyle{mnras}
\bibliography{references}

\bsp
\label{lastpage}
\end{document}